\documentclass[10pt,twocolumn,letterpaper]{article}
\usepackage{cvpr}              

\usepackage{graphicx}
\usepackage{amsmath}
\usepackage{amssymb}
\usepackage{booktabs}
\usepackage{algorithm}
\usepackage{algorithmic}
\usepackage{array}
\usepackage{xcolor}
\usepackage{url}
\usepackage[misc]{ifsym}

\usepackage[breaklinks,colorlinks]{hyperref}
\usepackage[accsupp]{axessibility} 

\usepackage[capitalize]{cleveref}
\crefname{section}{Sec. }{Secs. }
\Crefname{section}{Section}{Sections}
\Crefname{table}{Table}{Tables}
\crefname{table}{Tab. }{Tabs. }

\def\cvprPaperID{4051} 
\def\confName{CVPR}
\def\confYear{2022}

\def\xuef{\textcolor{black}}

\newcommand{\myparagraph}[1]{\vspace{1pt} \noindent \textbf{#1} }

\begin{document}

\title{Revisiting Document Image Dewarping by Grid Regularization}


\author{Xiangwei Jiang\textsuperscript{1,2}\thanks{Equal Contribution}, Rujiao Long\textsuperscript{3}\footnotemark[1], Nan Xue\textsuperscript{1}, Zhibo Yang\textsuperscript{3}, Cong Yao\textsuperscript{3}, Gui-Song Xia\textsuperscript{1,2}\thanks{Correspondence Author}\\ 
\textsuperscript{1}School of Computer Science, Wuhan University, Wuhan, China\\ 
\textsuperscript{2}LIESMARS, Wuhan University, Wuhan, China \qquad
\textsuperscript{3}Alibaba-Group, Hangzhou, China
}

\maketitle

\begin{abstract}
This paper addresses the problem of document image dewarping, which aims at eliminating the geometric distortion in document images for document digitization. Instead of designing a better neural network to approximate the optical flow fields between the inputs and outputs, we pursue the best readability by taking the text lines and the document boundaries into account from a constrained optimization perspective.  Specifically, our proposed method first learns the boundary points and the pixels in the text lines and then follows the most simple observation that the boundaries and text lines in both horizontal and vertical directions should be kept after dewarping to introduce a novel grid regularization scheme. To obtain the final forward mapping for dewarping, we solve an optimization problem with our proposed grid regularization.
The experiments comprehensively demonstrate that our proposed approach outperforms the prior arts by large margins in terms of readability (with the metrics of Character Errors Rate and the Edit Distance) while maintaining the best image quality on the publicly-available DocUNet benchmark.
\vspace{-1.5em} 
\end{abstract}
\section{Introduction}
\label{sec:intro}

The technologies of document digitization (DocDig) have largely facilitated our daily lives by transferring the information written in paper sheets from the physical world to electronic devices. Since the paper sheets are thin, fragile and easily-deformed, it is usually required to carefully capture document images of paper sheets with the scanners to avoid unexpected deformation of papers for digitization. Such a pipeline works well to some extent, however, it quickly loses efficiency when we are using handheld mobile devices, \eg, smartphones, for fast, painless but accurate document digitization. 
Therefore, the communities of computer vision and document analysis have been making efforts on getting rid of the restriction of DocDig by studying the problem of document image dewarping. 

\xuef{
There is a rich history for the problem of document image dewarping. In the early pioneering works~\cite{brown2001document,meng2014active,tian2011rectification}, this problem was formulated as surface reconstruction from different imaging configurations including multi-view images, binocular cameras as well as structured-lightning depth sensors, achieving accurate results in the lab environment while remaining issues for practical usage. Subsequently, the prior knowledge of paper-sheets in single-view images was extensively studied by detecting the boundaries~\cite{brown2006geometric}, text lines~\cite{kil2017robust,kim2015document}, structured lights~\cite{meng2014active}, and so on~\cite{ulges2004document} in the pre-deep-learning era. As those approaches depend on the detected prior knowledge, they are limited by the detection quality of those low-level visual cues, thus posing an issue of accuracy for the dewarping. 
}

\begin{figure}[t]
\centering
\begin{subfigure}[b]{1.0\linewidth}
\centering
    \begin{minipage}[b]{0.32\linewidth}
        \centering
        \includegraphics[width=1\linewidth]{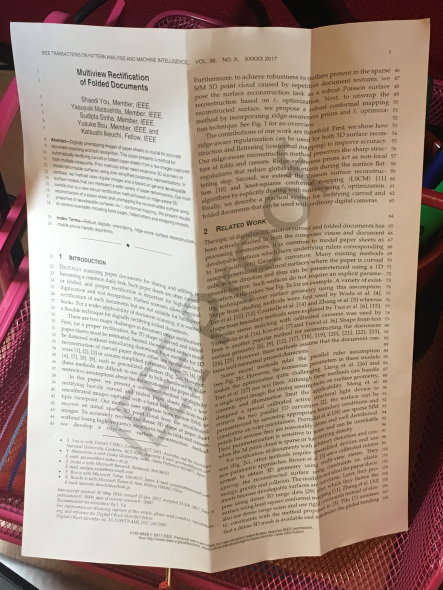} 
        \caption{Input Image}
    \end{minipage}   
    \begin{minipage}[b]{0.32\linewidth}
        \centering
        \includegraphics[width=1\linewidth]{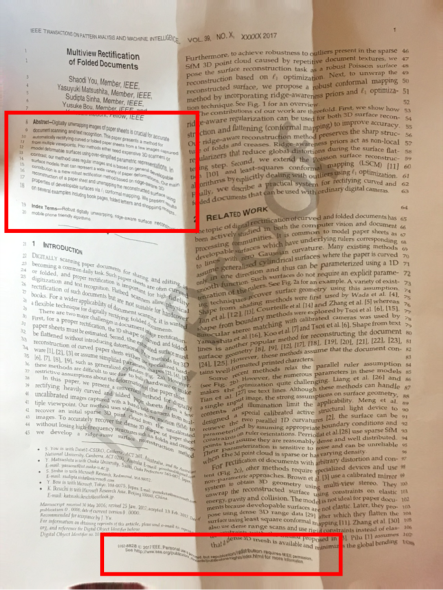} 
        \caption{DocUNet~\cite{Ma_2018_CVPR}}
    \end{minipage}  
    \begin{minipage}[b]{0.32\linewidth}
        \centering
        \includegraphics[width=1\linewidth]{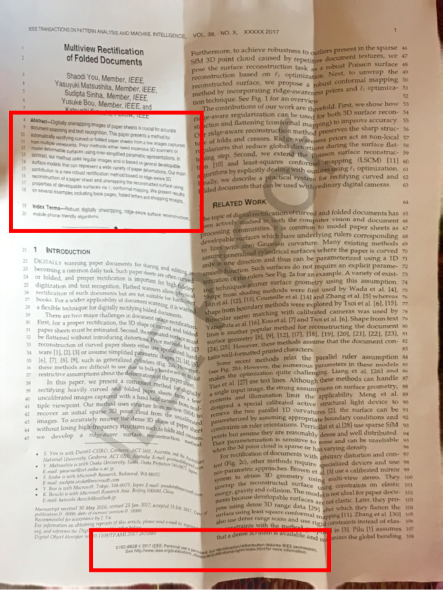} 
        \caption{DewarpNet~\cite{Das_2019_ICCV}}
    \end{minipage}  
    \begin{minipage}[b]{0.32\linewidth}
        \centering
        \includegraphics[width=1\linewidth]{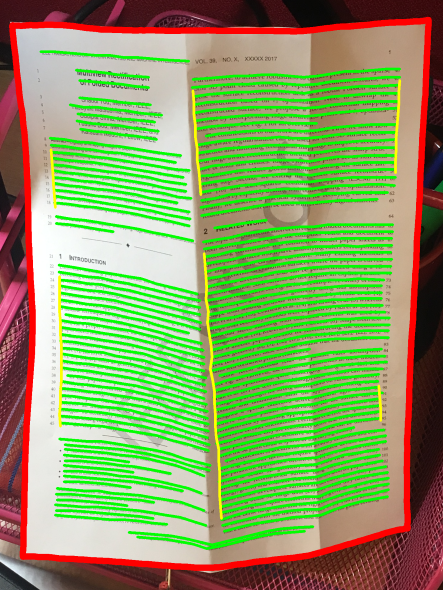} 
        \caption{Geometric Element}
    \end{minipage}   
    \begin{minipage}[b]{0.32\linewidth}
        \centering
        \includegraphics[width=1\linewidth]{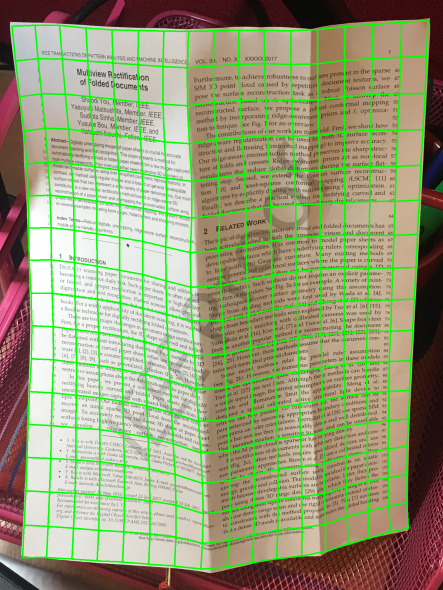} 
        \caption{Deformation Grid}
    \end{minipage}  
    \begin{minipage}[b]{0.32\linewidth}
        \centering
        \includegraphics[width=1\linewidth]{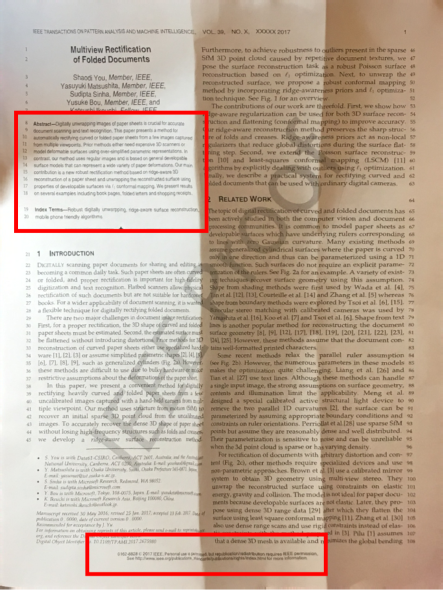} 
        \caption{Our Result}
    \end{minipage}  
    \vspace{-1em}
\end{subfigure}

\caption{Our method leverages the geometric information of text lines and document boundaries to generate a deformation grid which effectively reduces the error rate of character recognition.}
  \label{fig:fig1}
  \vspace{1em}
\end{figure}

\begin{figure}[!t]
  \centering
   \includegraphics[width=1\linewidth]{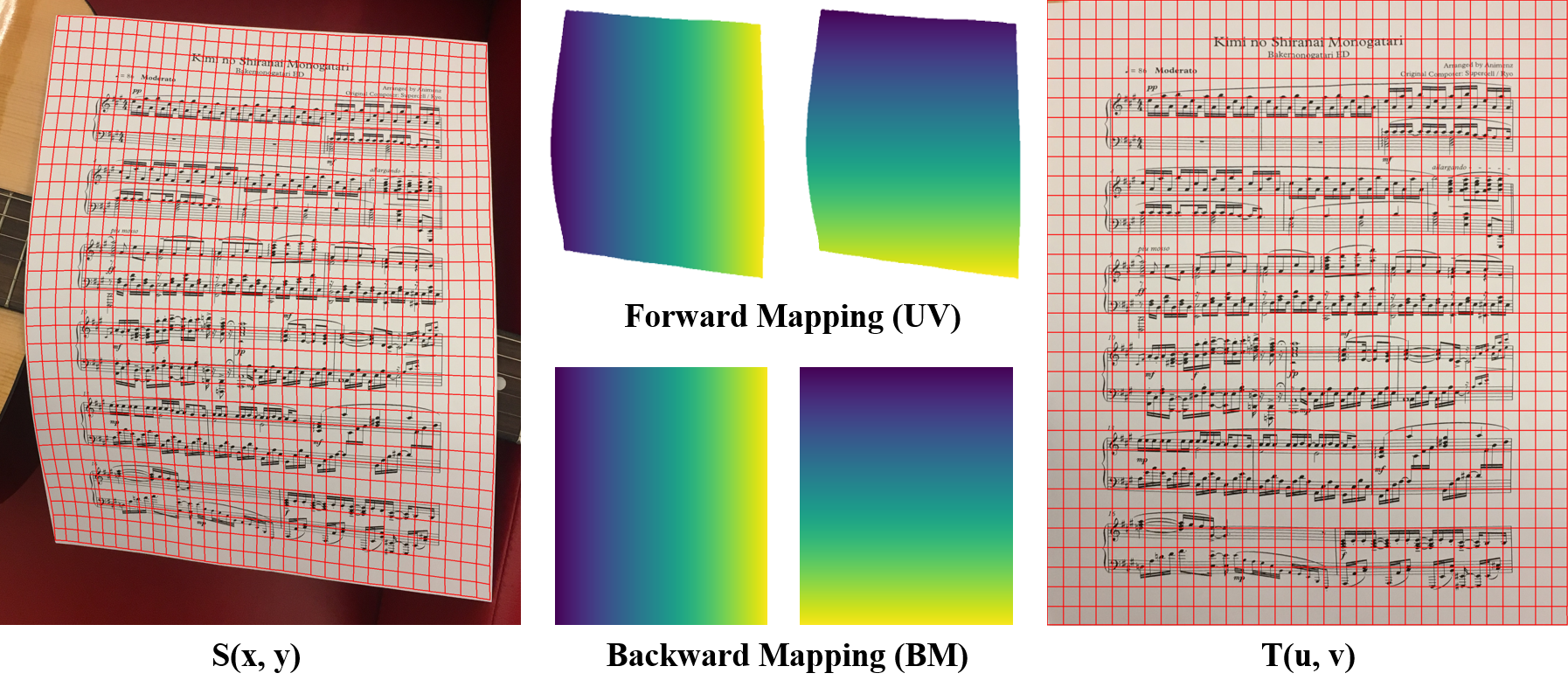}

   \caption{$S(x,y) = T(u,v)$: The equal sign means that pixel values of the corresponding position in different images are equal. Grids belonging to different images correspond to each other. $S \rightarrow T$ is backward mapping, while $T \rightarrow S$ is forward mapping.}
   \label{fig:DS}
   \vspace{1em}
\end{figure}

With the resurgence of deep learning in recent years, convolutional neural networks (ConvNets) have also been used for dewarping document image by learning the deformation field between the distorted input images and the expected flattened ones~\cite{Ma_2018_CVPR,Das_2019_ICCV,li2019document,CEARSE,xie2020dewarping,das2021end} under various supervision signals such as the synthetic data annotations~\cite{Ma_2018_CVPR}, the latent 3D shapes of documents~\cite{Das_2019_ICCV}, high-level semantics~\cite{CEARSE}, and so on~\cite{xie2020dewarping,das2021end}. 
These deep-learning based methods define the document image dewarping problem as a task of learning a 2-dimensional deformation field, which can move pixels from the original image $S$ (Source) to the geometrically rectified image $T$ (Target) as shown in Fig.~\ref{fig:DS}.
From the perspective of image quality, those deep-learning approaches significantly improved the dewarping accuracy. However, due to the low-frequency characteristic of neural networks, those approaches remain an issue of readability in the text regions of the output images.

In this paper, we revisit the deep-learning based approaches for document image dewarping and address the issue of readability by grid regularization.
Our proposed approach takes advantage of the inherent pattern of forward 2-dimensional deformation field\footnote{Our work relies on the assumption in the most common situations that the text lines and document boundaries are either horizontal or vertical in the rectified images.} and uses deep learning to detect some geometric information of document images, such as boundaries and text lines, to complete the task of dewarping from a single document image. 
More precisely, we first obtain boundary points and text lines through key-point detection and semantic segmentation respectively. Then, we estimate the vertical boundary of the text lines using their geometric properties. After that, this geometric information are discretized into 2-dimensional deformation field constraints, with which image deformation energy is minimized by integrating a grid regularization term. Finally, the rectified image is reconstructed, which conforms to our geometric prior characteristics (Fig. \ref{fig:fig1}).

{
Our contributions in this paper are as follows:
\begin{itemize}
\item We present a brand-new framework for document image rectification. The framework employs document boundaries and text lines to construct deformation field with a grid regularization term which is based on the traditional deformation model.
\item We systematically analyze the relative merits of regular terms used in image rectification approaches and come up with a novel grid generation method that better optimizes the quality of the deformation gird. 
\item The proposed approach achieves state-of-the-art performance on the DocUNet benchmark, demonstrating the effectiveness of the grid regularization term.
\end{itemize}                      
}

\section{Related Works}
\label{sec:relat}

\myparagraph{Parametric model based approaches.} consider a document image as a simple parametric surface and estimate the corresponding parameters by relying on the detected geometric features in the image. Cylindrical model~\cite{courteille2007shape,koo2009composition,liang2008geometric,cao2003cylindrical}, coons mesh ~\cite{farin1999discrete} and more complicated developable surface ~\cite{liang2005flattening,brown2004image,gumerov2004structure} have been always chosen as the base models. The involved geometric features include 2D curves such as text lines~\cite{lu2006document,stamatopoulos2010goal}, document boundary~\cite{cao2003rectifying}, as well as 3D curves obtained by structured light cameras~\cite{brown2001document,meng2014active}. Recently, Kim and Kil {\em et al.} ~\cite{kil2017robust}~\cite{kim2015document} combine parameter estimation with the camera model that has attracted much attention.
While with good performance on document images with simple geometric distortions, these parametric models are too easy to tackle complex geometric distortions appearing in real-world applications.

\myparagraph{3D reconstruction based methods.} often have two stages: estimating the 3D shape of the document image and predicting a 2D deformation field. Many ways can be used to obtain a 3D point-cloud surface~\cite{you2017multiview} such as with depth camera~\cite{ulges2004document,zhang2008improved}, binocular camera, laser scanner ~\cite{zhang2008improved}, multi view images~\cite{you2017multiview} and even with text lines~\cite{tian2011rectification}. Subsequently, a 3D surface can be reconstructed from the point clouds and the corresponding 2D deformation field~\cite {brown2007restoring} is then recovered by retrieving the surface parameters~\cite{meng2011metric,meng2018exploiting}. 
Obviously, the calculation of the 2D deformation field in this kind of methods heavily depends on the estimation accuracy of the 3D surfaces, \eg, the continuity of the 3D surface and the accuracy of bending. In real-world applications, these methods often quickly lose their efficiency when the lighting conditions and backgrounds are not well-designed.

\myparagraph{Deep-learning based approaches.} on document image dewarping can be dated back to the work in ~\cite{Ma_2018_CVPR}, which approached the problem by learning the 2D deformation field with a stacked U-Net. According to the paradigm of 3D reconstruction, Das {\em et al.} ~\cite{Das_2019_ICCV} add 3D shape information into the network to get a better estimation of the deformation field. While Markovitz {\em et al.} ~\cite{CEARSE} use the constraint that the text line is perpendicular to the text line boundary in 3D shape and texture mapping estimation of DewarpNet~\cite{Das_2019_ICCV} to predict the deformation field. Xie {\em et al.}~\cite{xie2020dewarping} improve forward optical flow estimation by using the similarity of adjacent optical flow. More recently~\cite{xie2021document}, they further proposed to use Encode structure to estimate fewer points with the Laplace grid constraint and obtain a better deformation field by traditional interpolation algorithm. Li {\em et al.}~\cite{li2019document} predict a series of small parts of the accurate deformation field by dividing the deformation field and get global result by integral constraints. Das {\em et al.} ~\cite{das2021end} integrate this thought into DewarpNet~\cite{Das_2019_ICCV}, resulting in an end-to-end model.  
Although these approaches~\cite{ramanna2019document,liu2020geometric,feng2021doctr} are designed to rectify the whole image, they tend to ignore the detailed text content which greatly affects the readability of the rectified document.
\vspace{-1em}

\section{A Revisit to Document Image Dewarping}

Let $S:\Omega_S \mapsto \mathbb{R}^3$ be a RGB document image suffered from some geometric distortions on the image domain $\Omega_S \subset \mathbb{R}^2$ (\ie, image grid), the ultimate goal of document image dewarping is to pursue a geometric transformation 
$$
\phi: \Omega_S \mapsto \Omega_T
$$ 
for reconstructing a new RGB image $T: \Omega_T \mapsto \mathbb{R}^3$ such that there is no geometric distortion on the image domain $\Omega_T \subset \mathbb{R}^2$, meaning that $\Omega_T$ is flattened.

As being reviewed in Section~\ref{sec:relat}, traditional methods are apt to recover such geometric transformation $\phi$ either by relying on some paramedic models with geometric constraints, or by using to calculate the 2D deformation field with the help of 3D surface information reconstructed from the document images. While being simple, these methods often lose their efficiency when the geometric distortions appeared in the source image $S$ are complex, \eg the cases discussion in this paper.

Instead of estimating $\phi$ from a single document image, and benefiting from the strong capability of neural networks to approximate nonlinear functions, deep-learning based methods propose to learn such geometric transformation $\phi$ from a large set of annotated document images with their corresponding rectified version being given.   

In what follows, we revisit the problem of document image dewarping, including both traditional methods and deep learning based ones, from the point view of grid regularization.
First of all, we recall the traditional interpolation model~\cite{tsoi2007multi}, an elementary procedure that is widely used in DocDig.  
\vspace{-0.5em}

\subsection{Transfinite Interpolation} 
 
Given the four boundaries of a document image, the transfinite interpolation model~\cite{tsoi2007multi} can dewarp the document image by using parametric curves.

Denote $\mathbf{c}_{1}(t),\mathbf{c}_{2}(t),\mathbf{c}_{3}(t),\mathbf{c}_{4}(t)$ as the four curves of boundaries parameterized by $t$, 
and $\mathbf{c}(u,v)$ being the grid interpolation from the distorted domain $\Omega_S$ to the dewarping $\Omega_T$ with $u,v$ are the surface parameters, then the transfinite interpolation reads,
\begin{equation}
\begin{aligned}
\small
    \mathbf{c}(u, v) & =(1-u, u)
    \left(\begin{array}{l}\mathbf{c}_{4}(v) \\
    \mathbf{c}_{2}(v)\end{array}\right)
    +\left(1-v, v\right)
    \left(
    \begin{array}{l}\mathbf{c}_{1}(u) \\
    \mathbf{c}_{3}(u)\end{array}\right) \\
    &-\left(1-u, u\right)\left(\begin{array}{ll}
    \mathbf{c}_{1}(0) & \mathbf{c}_{2}(0) \\
    \mathbf{c}_{3}(1) & \mathbf{c}_{4}(1)
    \end{array}
    \right)
    \left(\begin{array}{c}
    1-v\\ v \end{array}\right).
\label{eq:tfi}
\end{aligned}
\end{equation}
The first two terms correspond to ruled surfaces and the last is a correction term which is a projection transformation. 
Ideally, the output grid points in $\Omega_T$ are uniform.

\subsection{Significance of Uniformly Deformation Grid}

Actually, there are strong geometric priors for document image dewarping. For instance, the pixel coordinates in the rectified image should be uniformly distributed on the grid. However, the deep-learning based methods heavily depend on the representation ability of the deep network and lack an efficient way to integrate such geometric constraints, which often results in a nonuniform output grid. 

Table~\ref{tab:Related method} summarizes the methods that integrate grid regularization into the document image dewarping problem since DocUNet~\cite{Ma_2018_CVPR}. One can see that the use of grid regularization has brought significant improvements to the problem in various methods. This motivates us to define a simple task to verify the validity of the regularization term.

\begin{table}[!t]
\scriptsize
  \centering
    \caption{Deep learning methods concerning regularization term.}
    \vspace{-3mm}
  \begin{tabular}{lccc}
    \toprule
    \textbf{Method} &  \textbf{Regularization Term}  & \textbf{MS-SSIM} $\uparrow$  & \textbf{LD} $\downarrow$\\
    \midrule
    DocUNet~\cite{Ma_2018_CVPR} & None   & 0.4389& 10.90\\
    DewarpNet~\cite{Das_2019_ICCV} & Checkerboard Reconstruction & 0.4692 & 8.98 \\
    Flow Estimation ~\cite{xie2020dewarping}&Similarity of Adjacent Flow & 0.4361&\textbf{8.50}\\
    Xie {\em et al.} ~\cite{xie2021document}& Laplacian Mesh & 0.4769&9.03\\
    Li {\em et al.} ~\cite{li2019document} & Image Splitting  &-  &-\\
    Das {\em et al.} ~\cite{das2021end} & Image Splitting and End-to-End& \textbf{0.4879}&9.23\\
    \bottomrule
  \end{tabular}
  \label{tab:Related method}
  \vspace{-1mm}
\end{table}

\begin{table}[!t]
\scriptsize
  \centering
    \caption{Traditional Interpolation with Boundary: The B in parentheses represents the boundary of the deformation field. We adopt DocUNet~\cite{Ma_2018_CVPR} and DewarpNet~\cite{Das_2019_ICCV} pretraining model to obtain the results in the table. At the same time, we utilize traditional interpolation algorithms (TFI, TPS) to obtain other results by using the boundary points of the deformation field from deep models.}
    \vspace{-3mm}
  \begin{tabular}{lcccc}
    \toprule
    \textbf{Method} &  \textbf{CER}($std$) $\downarrow$ & \textbf{ED} $\downarrow$ &\textbf{MS-SSIM} $\uparrow$ & \textbf{LD} $\downarrow$\\
    \midrule
    DocUNet~\cite{Ma_2018_CVPR} & 0.4872 (0.182) & 2051.84 & \textbf{0.4332} & 12.59	\\
    DocUNet(B)+TFI & 0.4537 (0.170) & 	1852.66 & 0.4240	& 12.85\\
    DocUNet(B)+TPS & \textbf{0.3861 (0.177)} & \textbf{1613.36} & 0.4133 & \textbf{11.92}  \\
    	
    \midrule
    DewarpNet~\cite{Das_2019_ICCV} & \textbf{0.3097 (0.193)} & \textbf{1360.51} & 0.4340	& 10.08	\\
    DewarpNet(B) +TFI & 0.3546 (0.203)	& 1488.25 & \textbf{0.4374} & 10.02	\\
    DewarpNet(B) +TPS &  0.3349 (0.194)	& 1488.00 & 0.4301 & \textbf{9.70} \\
    \bottomrule
  \end{tabular}
  \label{tab:bound}
  \vspace{2em}
\end{table}

We use DocUNet~\cite{Ma_2018_CVPR} and DewarpNet~\cite{Das_2019_ICCV} as baselines and compare two traditional interpolation algorithms, \ie, TransFinite Interpolation (TFI)~\cite{gordon1973transfinite} and Thin Plate Spline Interpolation (TPS)~\cite{bookstein1989principal}. The results are reported in Table~\ref{tab:bound}. 
For DocUNet~\cite{Ma_2018_CVPR}, which has no regularization constraint, a simple grid regularization method can bring significant improvement, saying that the {\em Character Error Rate} (CER) is decreased by $10.1\%$. But for DewarpNet~\cite{Das_2019_ICCV}, which has already been constrained by grid regularization, it is difficult for us to obtain good results based on the existing boundary conditions. Therefore,
it is of great interest to design a model that can be compatible with more geometric information and enable us to integrate grid regularization constraints.

\subsection{A Unified Document Image Dewarping Model}
To motivate the following studies, we first re-examine the dewarping function from the perspective of energy minimization. We approximate the solution by minimizing the transformation energy as in Eqn.~\eqref{eq:deformation-model1}. 
\vspace{-0.5em}
\begin{equation}
    \begin{split}
        \varepsilon = \varepsilon_{	\phi} &+ \lambda \varepsilon_{d},\\
        \phi : \, (x,y) &\mapsto  (u,v),\\
             (x,y)\in\Omega_S, \, &(u,v)\in\Omega_T.
    \end{split}
    \label{eq:deformation-model1}
\end{equation}

The $\varepsilon$ represents the total energy of the dewarping system; $\phi$ represents the deformation which transforms the source image $S$ into the target image $T$, that is $\phi(x,y) = (u,v)$, with $(x,y) \in \Omega_S$ being the coordinate point in the domain of $S$, $(u,v) \in \Omega_T$ being the corresponding coordinate point in the domain of $T$. $\varepsilon_{\phi}$, $\varepsilon_{d}$ represent the data penalty energy (usually points displacement) and surface distortion energy, respectively. $\lambda$ is a hyper parameter to balance the energies between the data penalty and the distortion. The solution with the smallest total energy is the image deformation we expected. 

We can use this uniform image dewarping model to summarize some previous methods including both traditional methods and deep learning methods.

Let $\bf p$ and $\bf q$ represent the source coordinate points and the target coordinate points, respectively. $\varepsilon_{\phi}$ is the control points displacement loss according to Eqn.~\eqref{eq:tps1}.
\vspace{-0.5em}
\begin{equation}
  \varepsilon_{\phi}= \sum_{i=1}^{N} \lVert \phi({\bf p_{i}})- {\bf q_{i}}\rVert_{2}^{2}.
  \label{eq:tps1}
\end{equation}

For the traditional methods, if taking $\varepsilon_{d}= \iint_{\mathbb{R}^{2}}( \phi_{xx}^{2}+ \phi_{yy}^{2})dxdy$ and $N = 4$, we will get the projection transformation. And the transfinite interpolation (TFI) also fits this form, which regards boundary points as control points and uses the same $\varepsilon_{d}$ as the projection transformation. For the thin plate spline (TPS) interpolation, $\varepsilon_{\phi}$ characterize the displacement of boundary points while $\varepsilon_d$ is an integral to characterize the degree of deformation, as in
\begin{equation}
  \varepsilon_{d}= \iint_{\mathbb{R}^{2}}(\frac{\partial^{2} \phi }{\partial x^{2}})^{2}+2(\frac{\partial^{2} \phi }{\partial x\partial y})^{2}+(\frac{\partial^{2} \phi }{\partial y^{2}})^{2}dxdy.
  \label{eq:tps2}
\end{equation}

In the deep-learning based methods, we add all points to points displacement energy $\varepsilon_{	\phi}$. As for the second item, DocUNet~\cite{Ma_2018_CVPR} has $\lambda = 0$. While DewarpNet~\cite{Das_2019_ICCV} has a regularization term with binary classification using checkerboard. The flow estimation~\cite{xie2020dewarping} method has a constraint on local adjacency flow. And Xie {\em et al.}~\cite{xie2021document} achieves accurate grid prediction by reducing the number ($N$) of control points.

\section{The Proposed Method}
\begin{figure*}[!t]
  \centering
   \includegraphics[width=.92\linewidth]{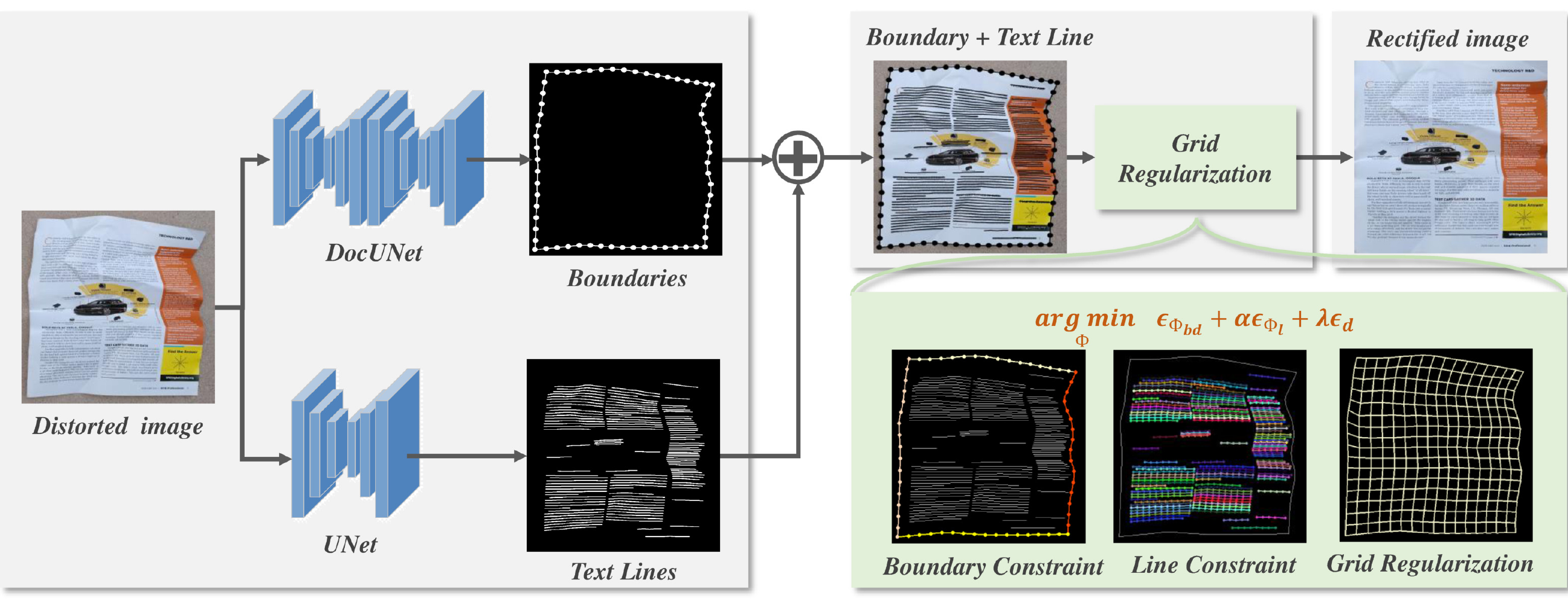}
   \vspace{-2mm}
   \caption{The pipeline of the proposed method. Taking a document image as input, DocUNet detects the backward mapping(BM) of document image boundaries by regression and UNet detects text lines by segmentation. The obtained geometric elements are discretized into points and fed into the grid regularization module. The proposed regularization module takes the boundary constraint, text line constraint and the grid regularization term as optimization conditions to calculate a uniform forward map (UV map).}
  \label{fig:pipeline}
\end{figure*}

Similar to DocUNet~\cite{Ma_2018_CVPR}, we approach the document image dewarping problem by calculating a forward 2D deformation field. In our model (Fig.~\ref{fig:pipeline}), we first detect geometric elements such as the boundaries of the document and text lines in the document images and discretize them into points. Next, we minimize an objective function incorporating the grid regularization to obtain an optimal deformation field. Compared with the previous methods, our model utilizes explicit geometry priors to dewarp the distortions in document images. 

\subsection{Grid Regularization with Text Lines}

In this subsection, we elaborate on how to use the unified document image deformation model to generate a deformation grid for our problem.  We discretize the geometric conditions of lines into the energy form of points and use a regularization term to constrain the deformation grid.  

\myparagraph{Text lines and vertical lines.} We use a common semantic segmentation network, \ie UNet, to extract text lines from document images, and obtain the boundaries of text lines by Algorithm~\ref{alg:Framwork}.
\vspace{-1em}
\begin{algorithm}[!htb] 
\renewcommand{\algorithmicrequire}{\textbf{Input:}}
\renewcommand{\algorithmicensure}{\textbf{Output:}}
\caption{Detecting the Boundary of Text Lines} 
\label{alg:Framwork} 
\begin{algorithmic}[1] 
\REQUIRE ~~\\ 
Text lines of the document image;\\
Hyper-parameter: $w$ = 15 pixels, $h$ = 15 pixels, $\theta = \arctan (0.45)$;\\
\ENSURE ~~\\ 
The boundaries of text lines;

\STATE Let $A$ is the set of left endpoints $d(x,y)$ of text lines,  and $B$ is the set of right endpoints.  

\STATE Calculate the vertical direction $g$ of each endpoint, which is determined by the average gradient of the three adjacent control points to itself. 

\STATE For each $d(x,y)$$\in$$A$, we search for another point $e \in A$ which is nearest to point $d$ in the region of $\left[x-w:x+w,y-h:y\right]$ and the line gradient between $d$ and $e$ with the condition that $g$ is less than $\theta$. Then connect $d$ and $e$. 

\STATE We do the same thing for $B$, the connecting lines form the connected component $M_1$. \\

\STATE Change the search area to $\left[x-w:x+w,y:y+h\right]$ to obtain the connected component $M_2$. \\ 
\STATE The final result of $M_{1}\cap M_{2}$ which can effectively avoid some error situations.  
\end{algorithmic}
\end{algorithm}

\begin{figure}[!t]
  \centering
  \vspace{-2mm}
  \includegraphics[width=0.32\linewidth, height=0.42\linewidth]{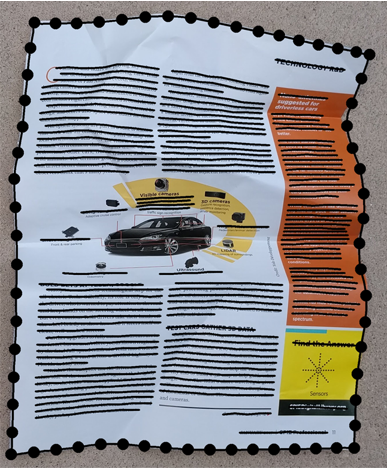}
  \includegraphics[width=0.32\linewidth, height=0.42\linewidth]{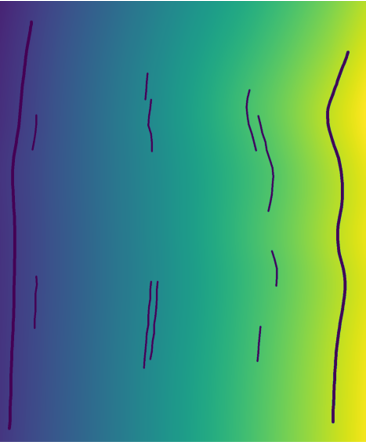}
  \includegraphics[width=0.32\linewidth, height=0.42\linewidth]{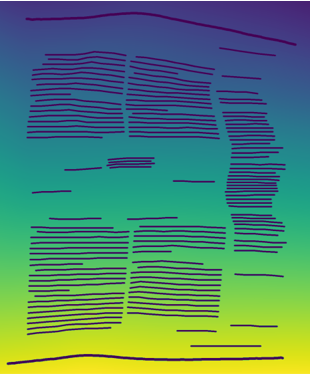}
 \vspace{-2mm}
   \caption{The $UV$ patterns that can be used to develop some equality constraints on $UV$ map.}
   \label{fig:uv}
  \vspace{1em}
\end{figure}

\myparagraph{$\bf UV$ patterns in source image.} As illustrated in Fig.~\ref{fig:uv}, from the visualizations of deformed grids, we find that the position of the boundaries has some characteristic in the forward mapping: Points on the left boundary have a $0$ value, and points on the right boundary has a value of 1 in the U map. And the position values increase uniformly from the left to the right in the $U$ map. Being the same as $U$ map, points in the 
$V$ map have the value of $0$'s or $1$'s for the top and the bottom boundary, respectively. Such rules have practical implications that we can make the boundary in $S$ go where it should be in $T$.
The points on the same vertical line share the same $u$ value, while the points on the same text line have the same $v$ value, which can guide the lines alongside the text line to go horizontal and the lines perpendicular to the text line to go vertical in $T$.
These rules can be used to develop some equality constraints on the $UV$ map.


\myparagraph{Establishment of optimization problem.} We first discretize lines into points. Therefore, we have boundary points, text line points and vertical line points. For convenience, we split $\phi(x,y)$ into two parts:
$$
  \forall (x,y) \in \Omega_S, \, \phi(x,y) = \left(\phi_{1}(x,y), \,\phi_{2}(x,y) \right) = (u,v).
  \label{eq:phi}
$$
The energy of the points on the document image boundary have the form in Eqn.~\eqref{eq:ep1}. 
\vspace{-0.5em}
\begin{align}
  \varepsilon_{\phi_{bd}}= \sum_{k=1}^{K} \lVert \phi_{i}(x_{k},y_{k})-j\rVert_{2}^{2},
  \label{eq:ep1}
\end{align}
where $K$ is the number of points in every boundary, with $bd \in \{top, left, bottom, right\}$. 
If we set $(i,j) = (0,0)$, then $\varepsilon_{\phi_{bd}}$ implies that the points on the left boundary of the document image in $S$ need to go to the left boundary in $T$. $(i,j) = (0,1),(1,0),(1,1)$ represent right boundary, top boundary and bottom boundary, respectively.

The points on the same text line in $S$ have the following energy form in Eqn.~\eqref{eq:ep2}. 
\begin{align}
  \varepsilon_{\phi_{l}}= \sum_{k=1}^{J-1} \lVert \phi_{i}(x_{k},y_{k})-\phi_{i}(x_{k+1},y_{k+1})\rVert_{2}^{2},
  \label{eq:ep2}
\end{align}
with $(x_{k},y_{k})$ and $\phi_{i}(x_{k+1},y_{k+1})$ being two adjacent points on the same line. $J$ is the number of points in a particular text line. The same thing for the vertical line. A horizontal text line has $i = 1$, and a vertical line has $i = 0$.

Inspired by regularization terms in TFI and TPS, we define our regularization terms by Eqn.~\eqref{eq:epd}. 
\begin{equation}
  \varepsilon_{d}= \iint_{\mathbb{R}^{2}}(\frac{\partial^{2} \phi }{\partial x^{2}}+\frac{\partial^{2} \phi }{\partial y^{2}})^{2}+\beta(\frac{\partial^{2} \phi }{\partial x\partial y})^{2}dxdy.
  \label{eq:epd}
\end{equation}
In order to make the optimization problem be compatible with the interpolation of dense control points, we merge $\phi_{xx}$ and $\phi_{yy}$ that guarantee the smoothness of the grid. $\phi_{xy}$ keeps the shape of the grid.  

So far, we can write down our optimization problem in Eqn.~\eqref{eq:opt}, by solving which we can obtain the grid deformation $\phi$ from $S$ to $T$.
\begin{equation}
  \begin{aligned}
  \arg \min_{\phi} \, & \sum_{k=1}^{4}\varepsilon_{\phi_{bd_{k}}} +\alpha\sum_{k=1}^{N_{1}}\varepsilon_{\phi_{l_{k}}} + \lambda \varepsilon_{d}.
  \end{aligned}
  \label{eq:opt}
\end{equation}
where $N_1$ is the number of lines, and $\alpha$, $\lambda$ are parameters to balance different energy terms.

\begin{figure}[!t]
  \centering
  \includegraphics[width=0.90\linewidth]{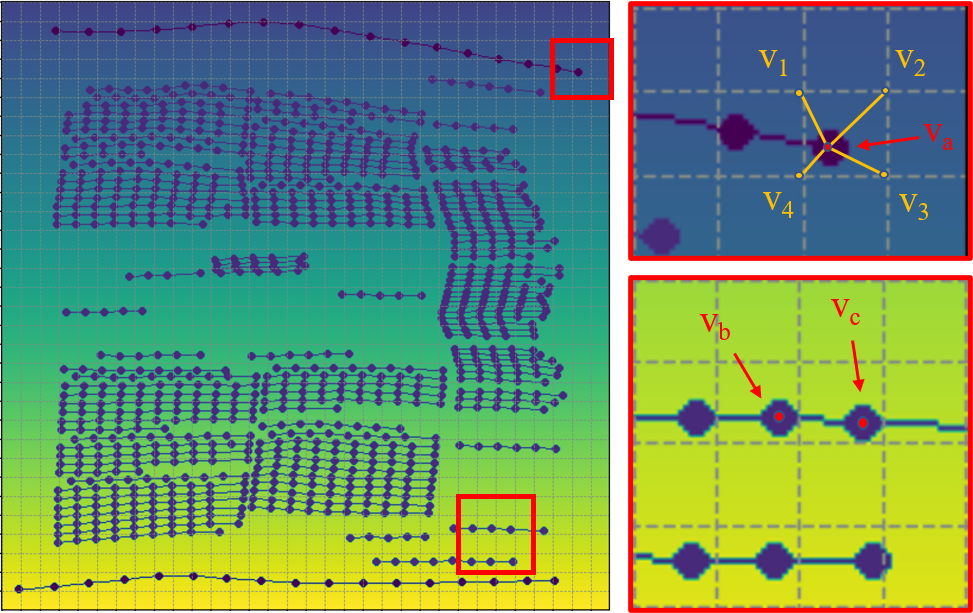}

   \caption{Calculation of V map: For every point a in V map: $v_{a}=\sum_{i=1}^{4} w_{i} \cdot v_{i}$, with $w_{i}$ being the coefficient of bilinear interpolation assigned to the four surrounding grid points $v_{i}$. For point a in the top boundary, we have $v_{a} = 0$. For point b, c in the same text line, we have $v_{b} - v_{c} = 0$. }
   \label{fig:uv_cal}
  \vspace{2mm}
\end{figure}

\myparagraph{Discretizing the optimization problem.}
To achieve a discrete solution of the transformation $\phi$, we need to discretize the data penalty energy and surface distortion energy. For a point $(x,y)$ on the boundary, $\phi(x,y)$ could be discrete to the nearest 4 grid points $\phi(x_{i},y_{i})$ by Eqn.~\eqref{eq:phi1}. 
\begin{equation}
\begin{aligned}
  \phi(x,y) = \sum_{i=1}^{4} w_{i}\,\cdot\,\phi(x_{i},y_{i}),
 \end{aligned}
  \label{eq:phi1}
\end{equation}
where $w_{i}$ is a bilinear interpolation coefficient, which has been used to apply known conditions to unknown grid points. The details of calculation are shown\footnote{Fig.~2 shows the \textit{groundtruth UV maps} while Fig.~5 illustrates the \textit{final UV predictions} that have the minimal deformation energy, therefore the background values are empty in Fig. 2.} in Fig. \ref{fig:uv_cal}. This form can also be applied to lines.

The distortion energy $\varepsilon_{d}$ can be discretized by,
\begin{equation}
\small
\begin{aligned}
  & \sum_{i,j} \Big(\phi[i+1,j] + \phi[i-1,j]
  + \phi[i,j+1] + \phi[i,j-1]-4\phi[i,j] \Big)^{2}\\
  & + \beta \sum_{i,j}  \Big(\phi[i+1,j+1] - \phi[i+1,j] - \phi[i,j+1]+\phi[i,j]\Big)^{2}. \\
\end{aligned}
\label{eq:phi2}
\nonumber
\end{equation}

\subsection{Implementation Details}

\myparagraph{Computing the document boundary constraints.} We obtain the document boundary by using DocUNet trained on the Doc3D dataset~\cite{Das_2019_ICCV}. As we focus on boundary information, we add some preprocessing to the boundary, such as color changes and flip operations. The image size for training is 128 $\times$ 128. And the train loss for DocUNet is defined as $L_1$ distance between the predicted boundary, \ie, $bd_{pred}$ with $bd \in \{top, bottom, left, right\}$ and the ground truth boundary of the deformation field $bd_{gt}$.

\myparagraph{Computing text-lines constraints.}
We train a UNet to detect text-line masks in document images with the help of PubLayNet~\cite{zhong2019publaynet} and the Scanned Script Dataset~\cite{Scanned2019}.
The image size for training is 512 $\times$ 512. The $UV$ labels from the Doc3D dataset~\cite{Das_2019_ICCV} are used to warp the image and text lines masks. 
Since the masks of the text lines are always much less than the background, we use an $L_2$ loss weighted by the pixel proportions~\cite{xue2019learning} to train the UNet model. 

\myparagraph{Grid regularization.} We discrete the line segments into points on a $512 \times 512$ image grid. The interval between points on the same text line is $16$ pixels and the interval between points on the same vertical line is $10$ pixels. By virtue of the geometry of these discrete points, one can redefine the data penalty term and image distortion term in the image distortion model accordingly. We set the hyper-parameter $n =128, \alpha =10, \lambda = 2, \beta = 20$ and we solve the optimization problem using Alternating Direction Method of Multipliers (ADMM)~\cite{diamond2016cvxpy} for Quadratic Programming (QP). When solving this optimization problem, due to the particularity of grid coordinates, we can optimize $\phi_{1}$ and $\phi_{2}$ separately and change the problem from a $n\times n \times 2$ dimension problem to two $n \times n$ dimension problems.

\myparagraph{Post-processing.} Once the forward map is obtained, we firstly generate the backward map (BM: UV $\mapsto \left[ 0,1 \right] \times \left[ 0,1 \right]$, LinearNDinterpolator) and then upsample the BM by a bilinear interpolation operation to obtain the high-resolution BMs. Then, for each pixel in the obtained BM, we sample the corresponding RGB value from the input image to yield the final results.

\section{Experiments and Analysis}
\label{sec:exper}

\begin{figure*}[!t]
\centering
\begin{subfigure}[b]{1.0\linewidth}
\centering
    \begin{minipage}[b]{0.16\linewidth}
        \centering
        \includegraphics[width=80pt]{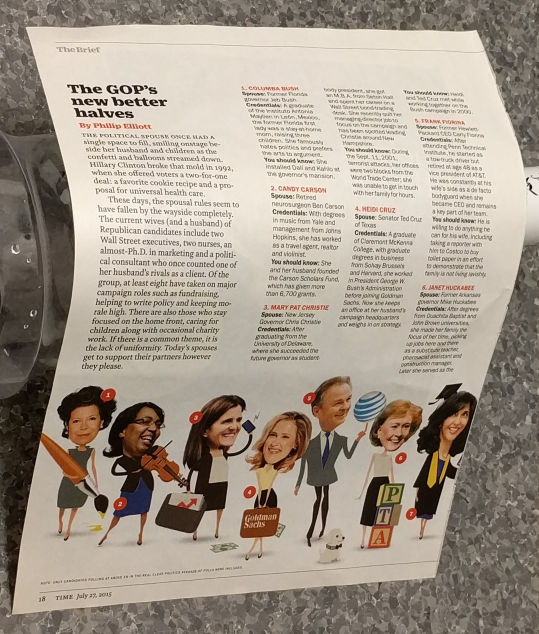}
        \includegraphics[width=80pt]{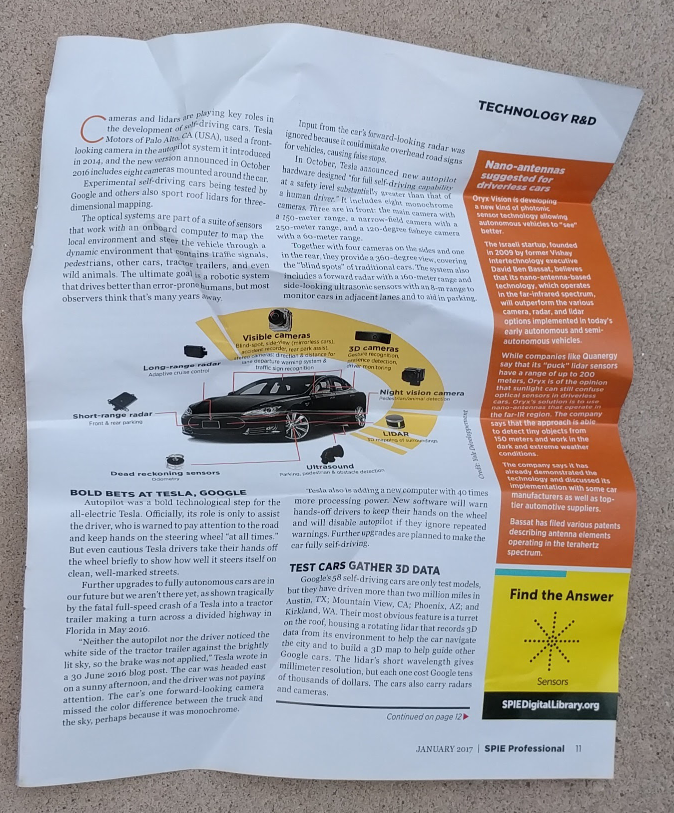}
        \includegraphics[width=80pt]{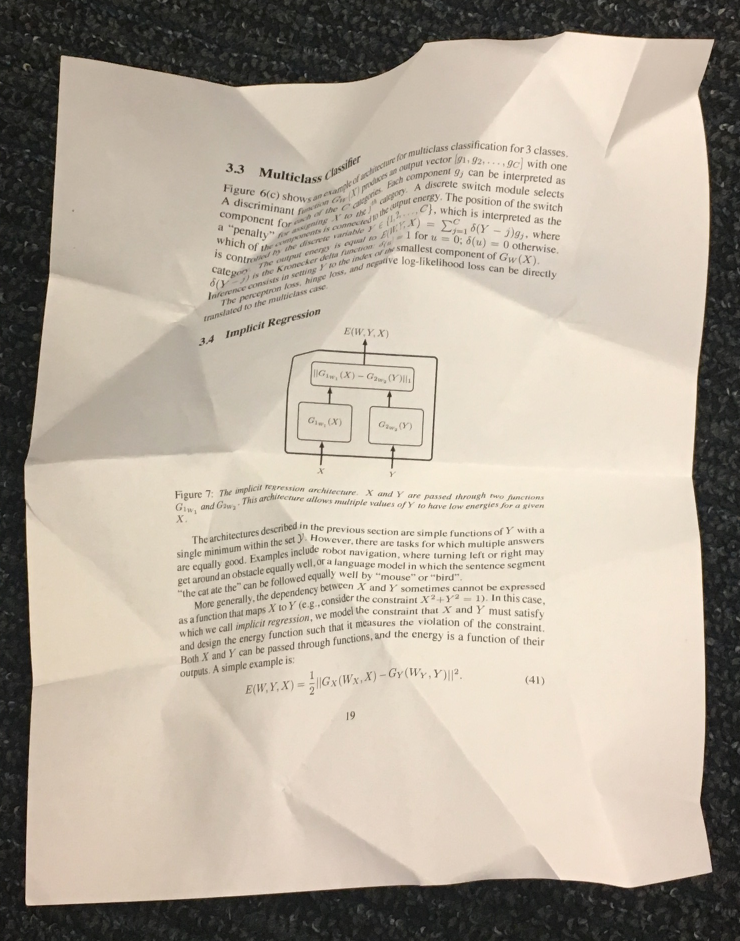}
        \includegraphics[width=80pt]{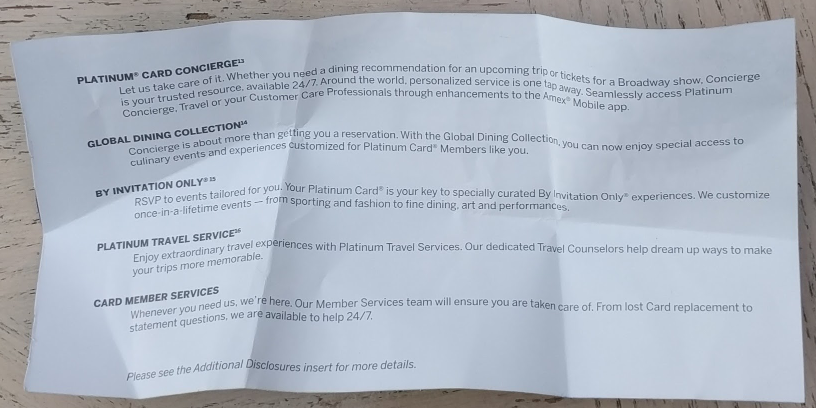}
        \caption{Original Image}
    \end{minipage}     
    \begin{minipage}[b]{0.16\linewidth}
        \centering
        \includegraphics[width=80pt]{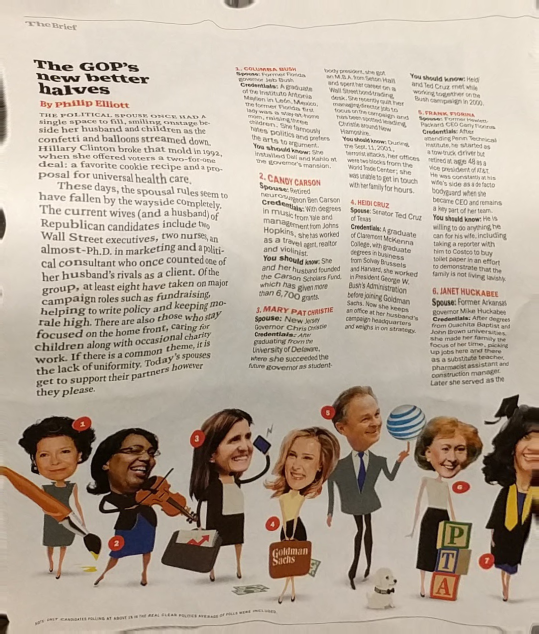}
        \includegraphics[width=80pt]{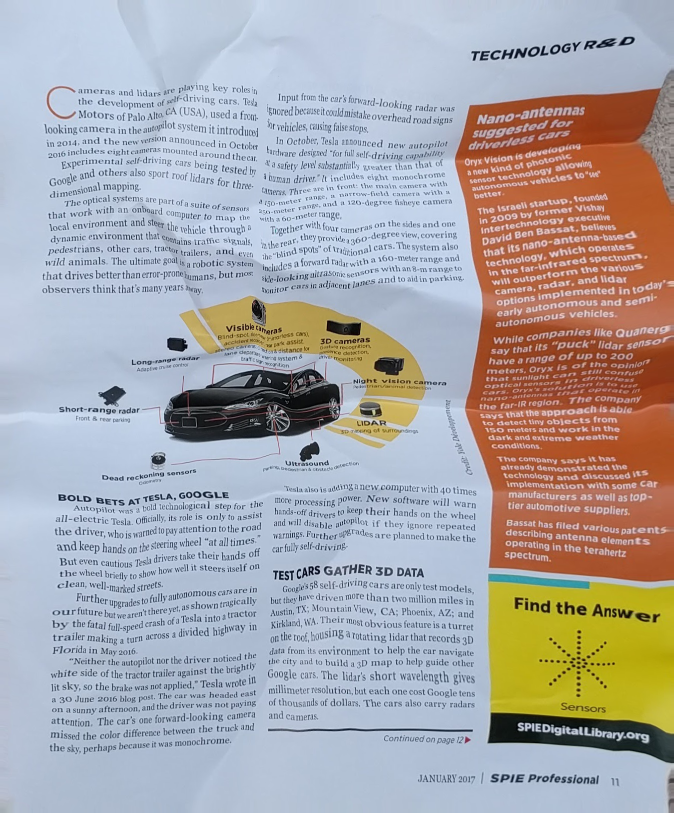}
        \includegraphics[width=80pt]{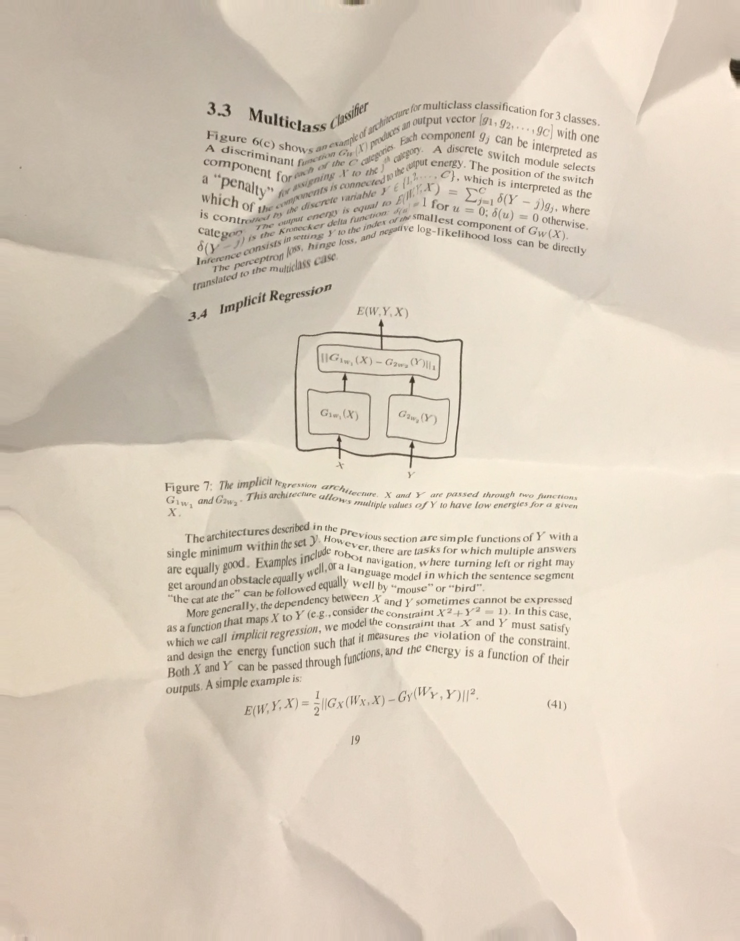}
        \includegraphics[width=80pt]{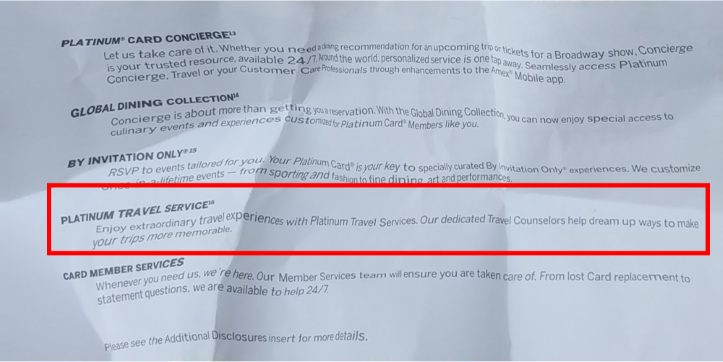}
        \caption{DocUNet}
    \end{minipage}  
    \begin{minipage}[b]{0.16\linewidth}
        \centering
        \includegraphics[width=80pt]{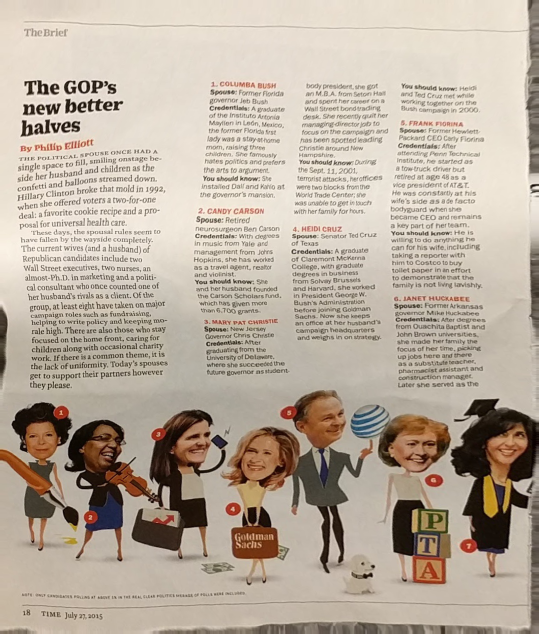}
        \includegraphics[width=80pt]{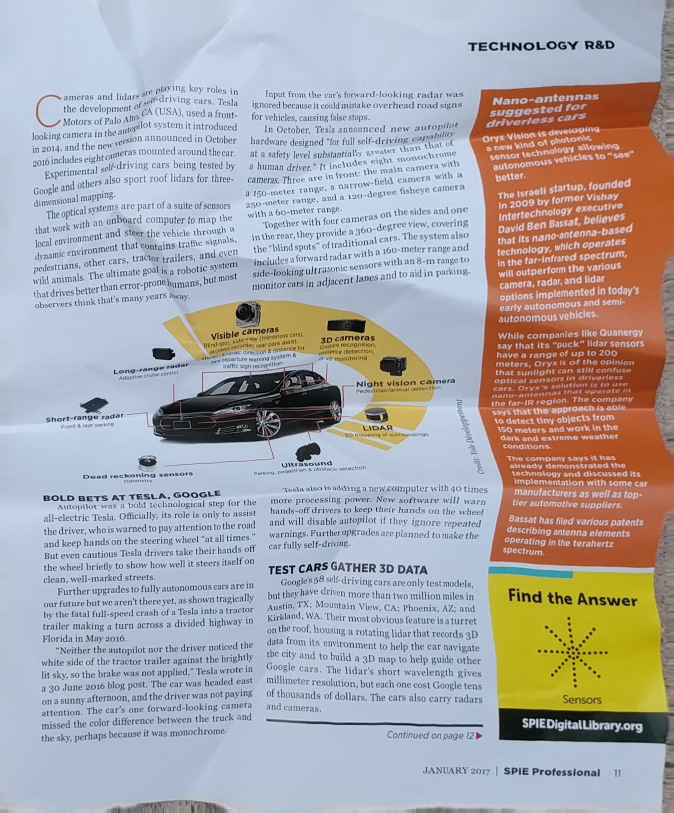}
        \includegraphics[width=80pt]{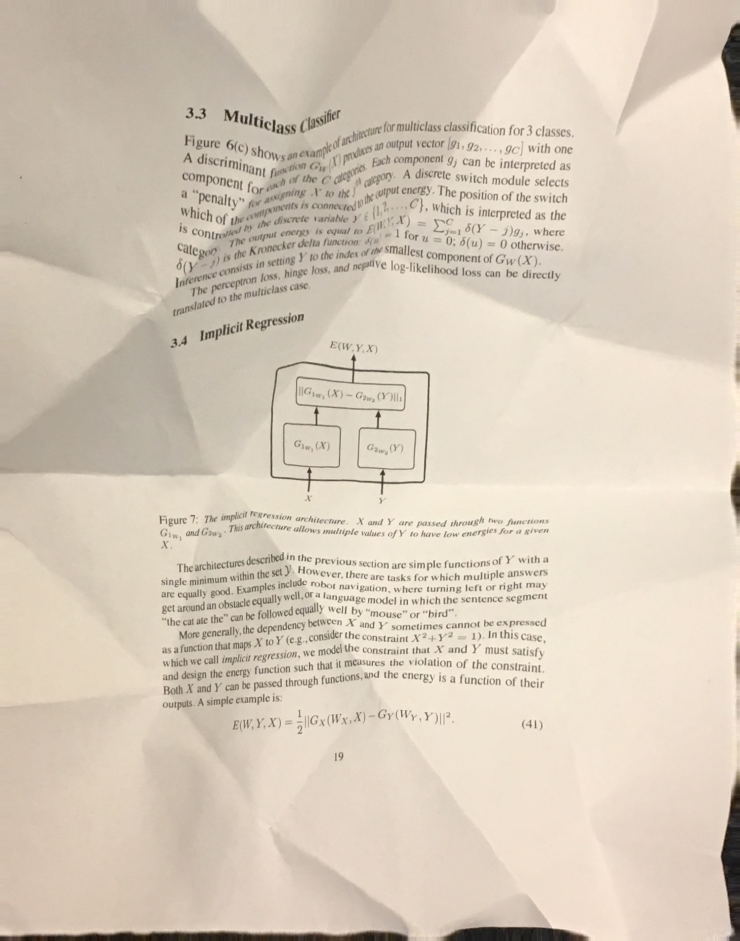}
        \includegraphics[width=80pt]{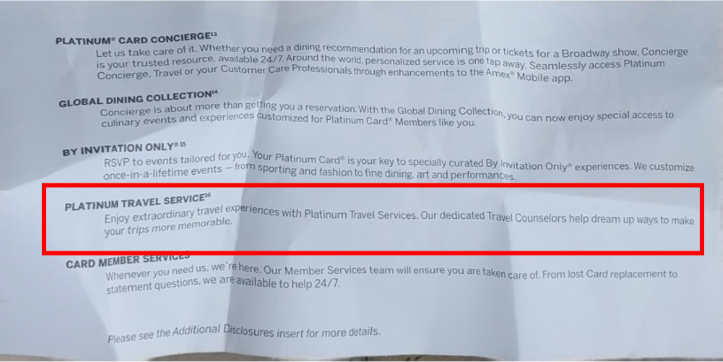}
        \caption{DewarpNet}
    \end{minipage} 
     \begin{minipage}[b]{0.16\linewidth}
        \centering
        \includegraphics[width=80pt]{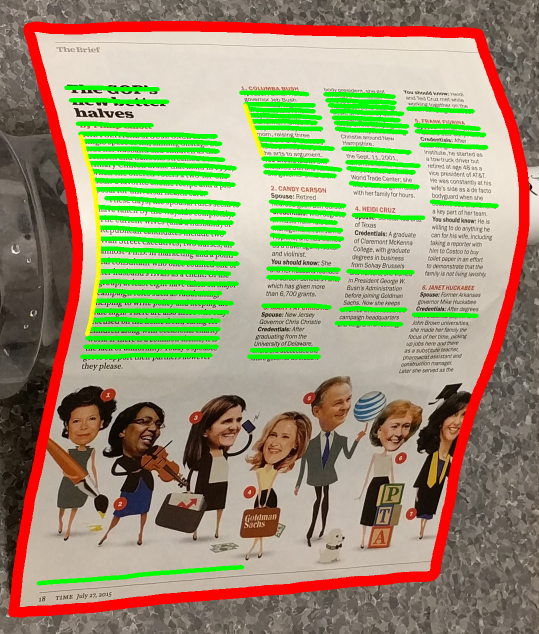}
        \includegraphics[width=80pt]{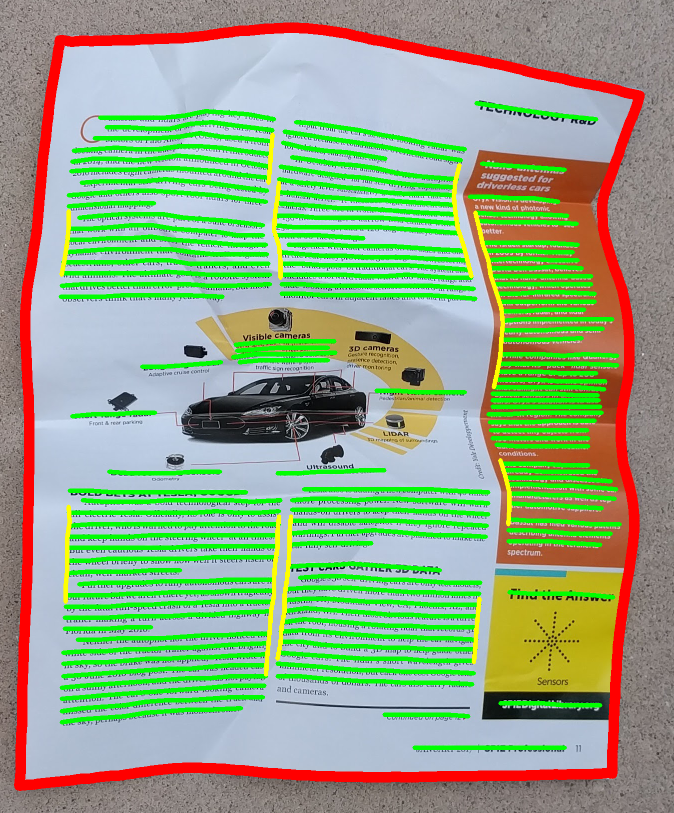}
        \includegraphics[width=80pt]{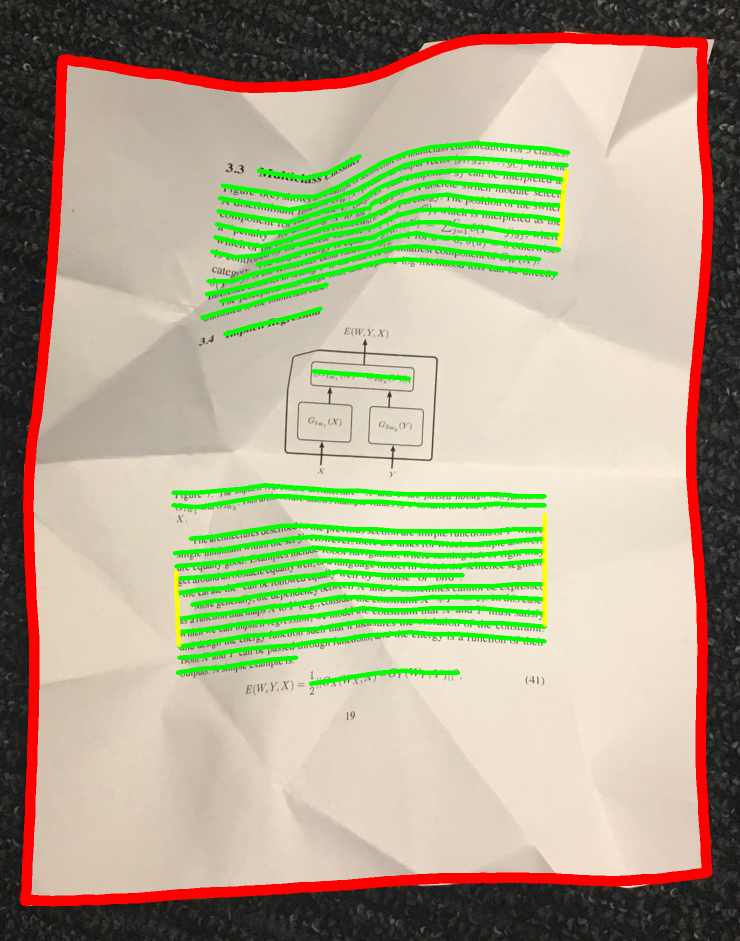}
        \includegraphics[width=80pt]{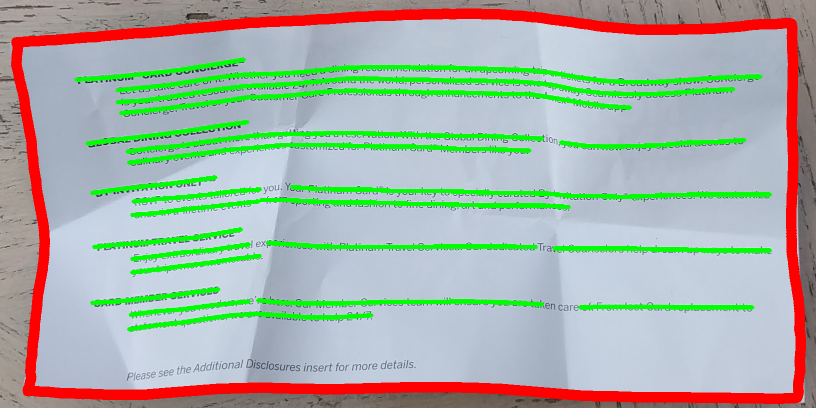}
        \caption{Geometric Elements}
    \end{minipage}  
    \begin{minipage}[b]{0.16\linewidth}
        \centering
        \includegraphics[width=80pt]{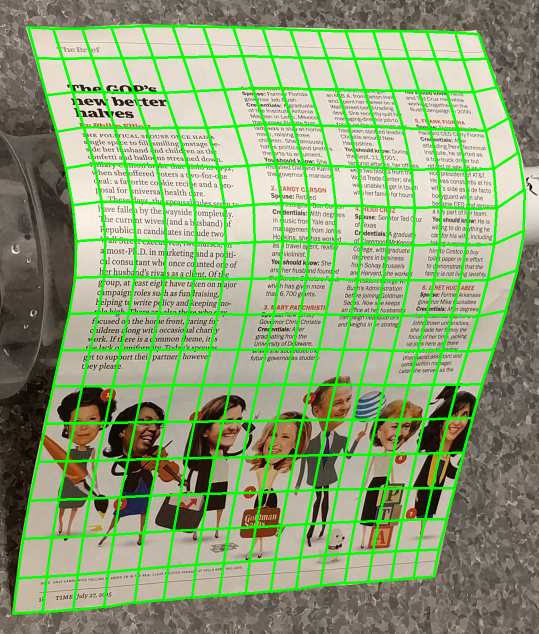}
        \includegraphics[width=80pt]{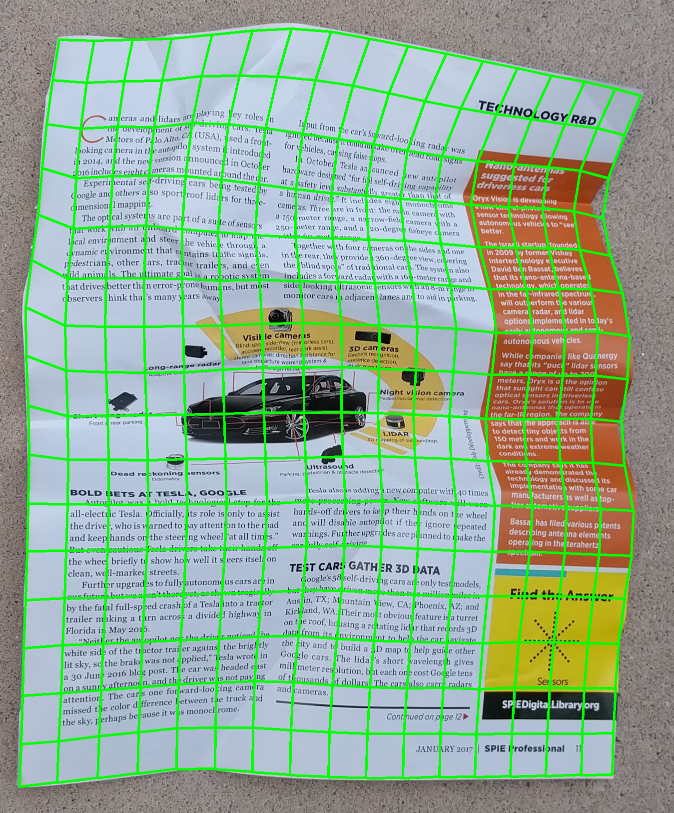}
        \includegraphics[width=80pt]{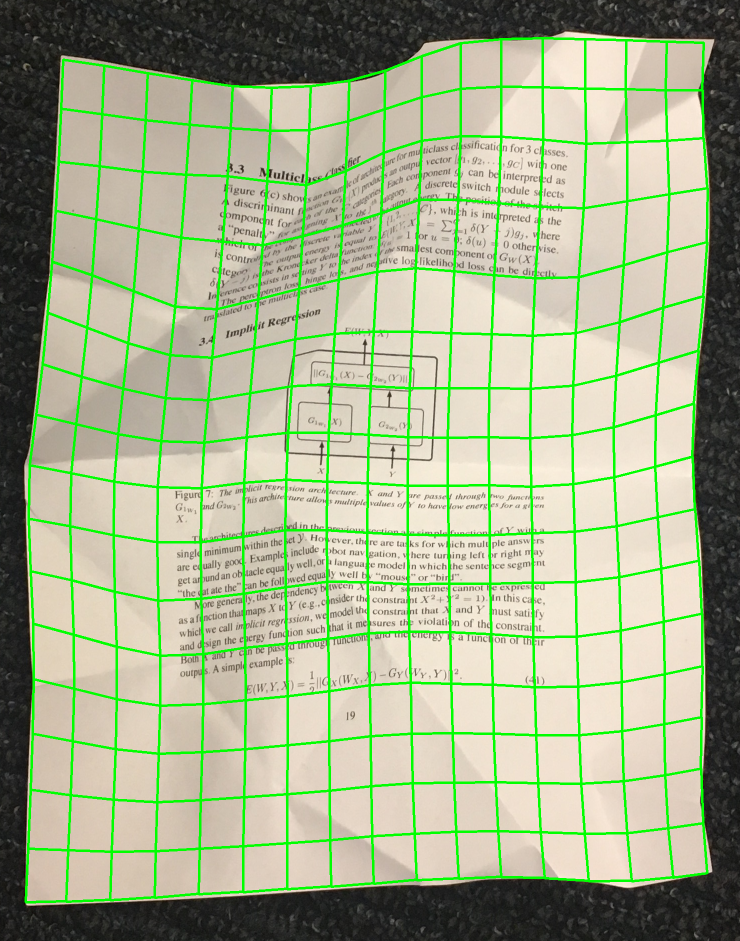}
        \includegraphics[width=80pt]{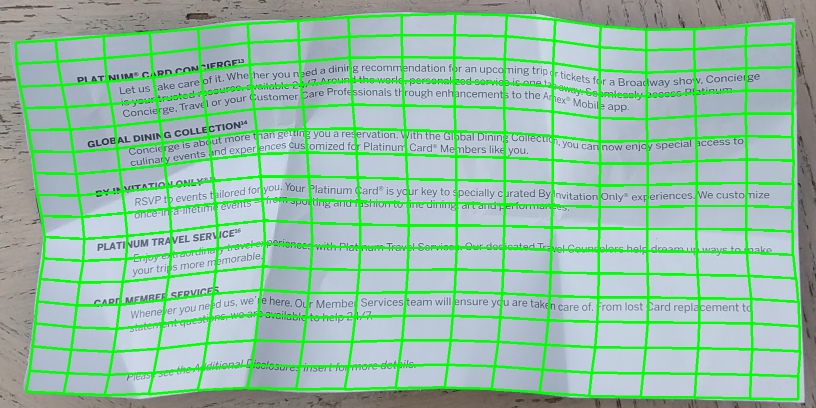}
        \caption{Deformation Grid}
    \end{minipage} 
    \begin{minipage}[b]{0.16\linewidth}
        \centering
        \includegraphics[width=80pt]{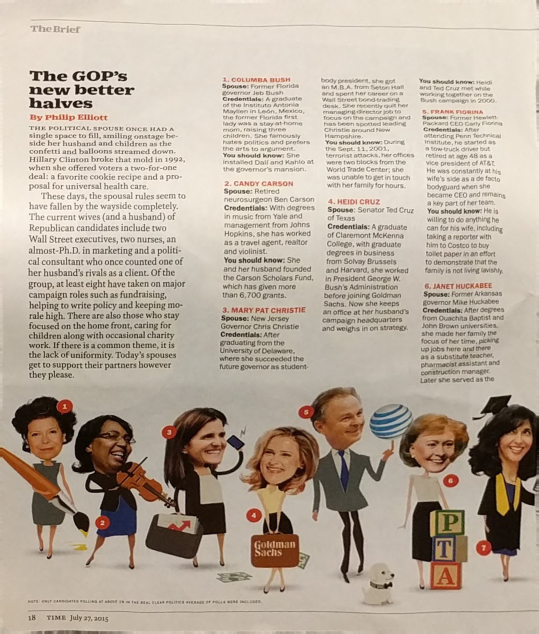}
        \includegraphics[width=80pt]{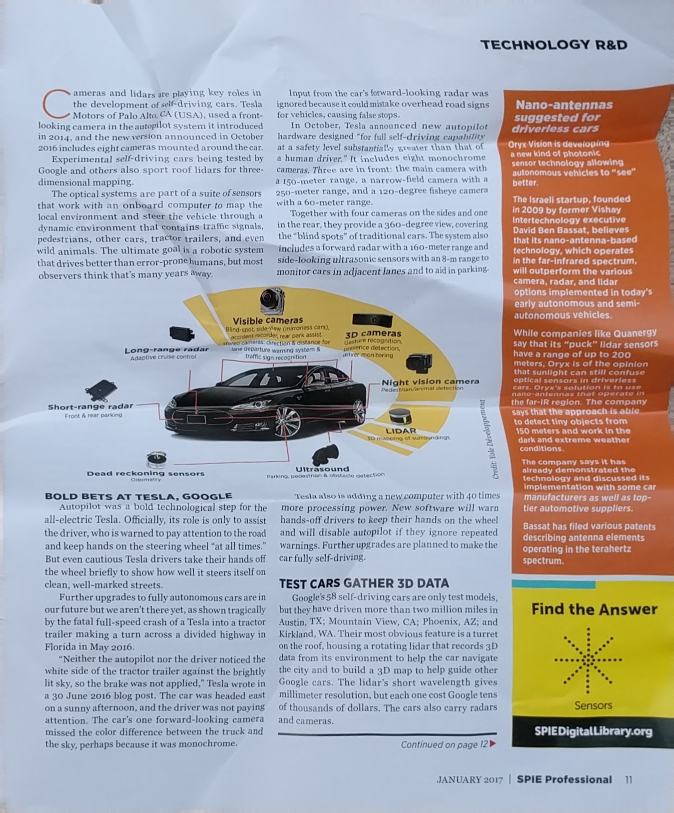}
        \includegraphics[width=80pt]{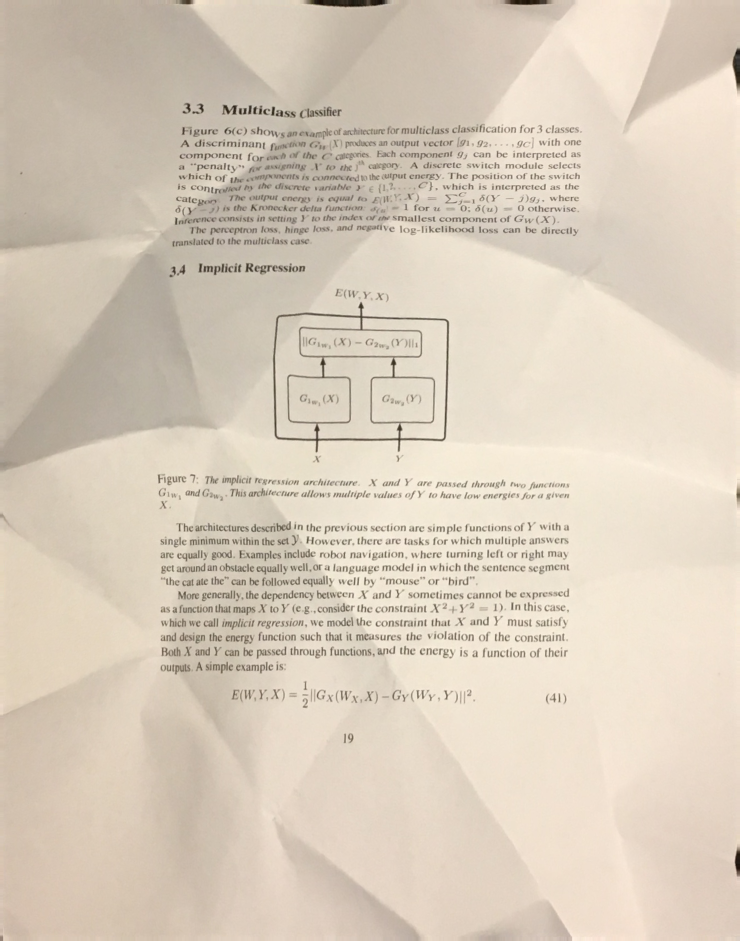} 
        \includegraphics[width=80pt]{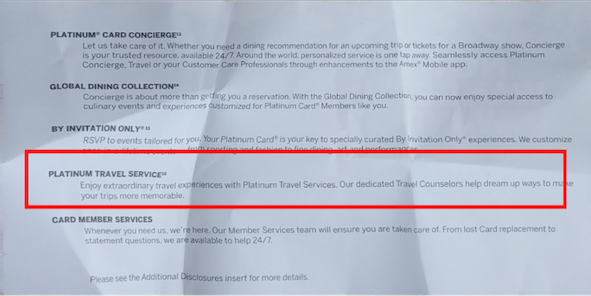}    
        \caption{Our Result}
    \end{minipage} 
  
\end{subfigure}
\caption{Qualitative comparison with previous methods: Our method uses text lines and vertical lines to guide the generation of deformation grid, and the rectified images are consistent with our expectations for Geometric Elements in document images. The green lines in the image of Geometric Elements are text lines and the yellow lines are the vertical lines extracted by our method.}
\vspace{-3mm}
\label{fig:result}

\end{figure*}

\subsection{Datasets and Evaluation Metrics}

\myparagraph{Dataset.} For experiments, we use the DocUNet~\cite{Ma_2018_CVPR} benchmark which has 130 images from real-world scenes and $50$ images with text labels~\cite{textlabels} which can help us analyze OCR performance.

\myparagraph{CER and ED:} We use CER (Character Error Rate) and ED (Edit Distance)~\cite{levenshtein1966binary} to evaluate a dewarping method. Specifically, for a rectified document image, CER computes the ratio of the unexpected deletions, insertions and substitutions among the reference string, while ED measures the dissimilarity between the OCR results from the rectified image and the ground truth labels.  For ED metric, we use PyTesseract (v0.3.8)~\cite{smith2007overview} for computation.

\myparagraph{MS-SSIM and LD:} 
For the rectified document images, we also leverage the image-based metrics of MS-SSIM (\ie, multi-scale SSIM)~\cite{wang2004image} and the Local Distortion (LD)~\cite{liu2010sift} metric to evaluate the results from global and local perspectives. In the implementation, we use the evaluation code provided by DocUNet for computation.

\subsection{Comparison with the State of the Arts}

\begin{table}[!t]
  \centering
  \scriptsize
   \caption{Quantitative comparsion of the proposed and previous methods on DocUNet benchmark. Standard deviation is reported in the parentheses. ``$\uparrow$'' indicates the higher the better and ``$\downarrow$'' means the opposite.}
   \vspace{-3mm}
  \begin{tabular}{lcccc}
    \toprule
    \textbf{Method} &  \textbf{CER}($std$) $\downarrow$ & \textbf{ED} $\downarrow$ &\textbf{MS-SSIM} $\uparrow$ & \textbf{LD} $\downarrow$\\
    \midrule
    DocUNet~\cite{Ma_2018_CVPR} & 0.3955 (0.272) & 1684.34  & 0.4389 & 10.90\\
    DewarpNet~\cite{Das_2019_ICCV} & 0.3136 (0.248) & 1288.60 & 0.4692 & 8.98\\  
    Flow Estimation ~\cite{xie2020dewarping} & 0.4472 (0.274) & 2000.04 & 0.4361 & \textbf{8.50} \\ 
    Xie {\em et al.} ~\cite{xie2021document} & -& - &0.4769 & 9.03\\ 
    Das {\em et al.} ~\cite{das2021end} & 0.3001 (0.14) & - & 0.4879 & 9.23\\
    Ours & \textbf{0.2068 (0.141)} & \textbf{896.48} & \textbf{0.4922} & 9.36\\ 
    \bottomrule
  \end{tabular}
  \label{tab:result}
    \vspace{1em}
\end{table}

We quantitatively and qualitatively compare our method with the S.O.T.A. deep learning methods~\cite{Ma_2018_CVPR,Das_2019_ICCV,xie2021document,li2019document,das2021end}. In those approaches, DocUNet~\cite{Ma_2018_CVPR} did not use any grid regularization while DewarpNet~\cite{Das_2019_ICCV} using the checkerboard reconstruction term for dewarping. 
For the flow-based approach, Xie {\em et al.}~\cite{xie2021document} utilize Laplacian grid to achieve a better performance. We also compared our method with the splitting-based approaches~\cite{li2019document,das2021end}.

\myparagraph{Quantitative Comparison.} 
As reported in Table~\ref{tab:result}, our proposed approach achieves the best performance for the metrics of CER, ED and MS-SSIM. Benefiting from the regularization formulation with text lines and the boundaries, our method reduces the CER by 9 points and ED by 30\%. For the image-based metrics of MS-SSIM and LD, our method obtains better structural similarity while maintaining a comparable result of the local distortion.

\begin{figure}[t]
  \centering
  \includegraphics[width=.85\linewidth]{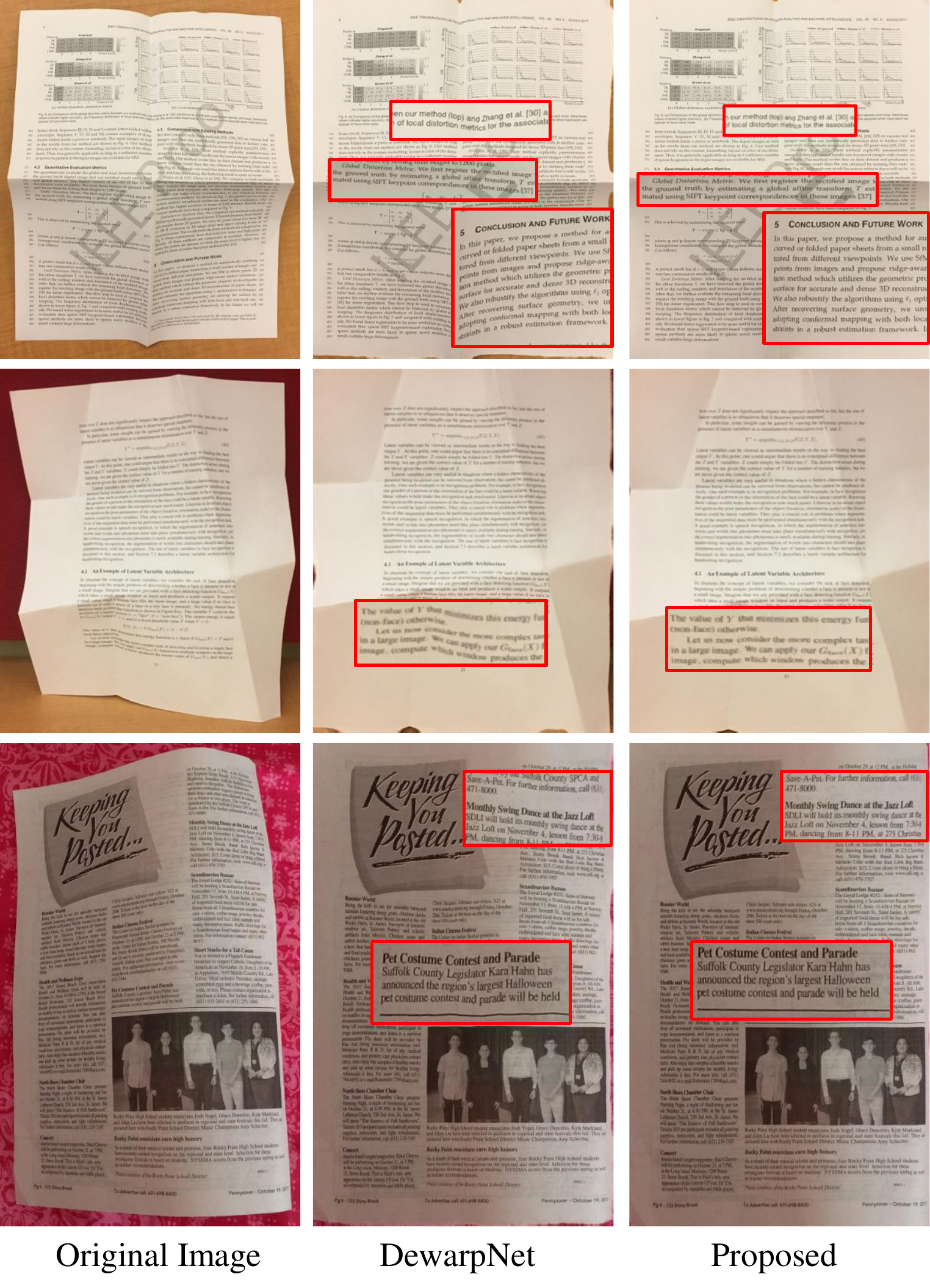}
  \vspace{-2mm}
  \caption{Local comparison of the proposed method with DewarpNet. The document image obtained by our proposed method has better readability. The red boxes show the enlarged details.}
  \label{fig:details}
\end{figure}

\myparagraph{Qualitative Comparison.} \cref{fig:result} shows the qualitative comparison with the prior arts. It is illustrated that our proposed method handles severe distortions very well. For the interior of the rectified images, our dewarping results are more smooth than DocUNet and DewarpNet as we did not use the predicted the deformation field in the interior regions. Alternatively, we obtain the deformation field in the internal regions via optimization with our proposed regularization term. Furthermore, we show the local comparison between our approach and DewarpNet in~\cref{fig:details}. 





\subsection{Ablation Study}
\begin{table}
  \scriptsize
  \centering
  \caption{Ablation experiment on different the grid regularization terms. We replace the proposed grid regularization with the traditional interpolation algorithm (TFI, TPS) for comparison. In the case of the same boundary control points, our proposed method can get slightly better results than the traditional interpolation algorithm in the image quality evaluation index.}
  \vspace{-2mm}
  \begin{tabular}{lcccc}
    \toprule
    \textbf{Method} &  \textbf{CER}($std$) $\downarrow$ & \textbf{ED} $\downarrow$ &\textbf{MS-SSIM} $\uparrow$ & \textbf{LD} $\downarrow$\\
    \midrule
    DocUNet & 0.3955 (0.272) & 1684.34 & 0.4389 & 10.90\\
    Boundary+TFI & 0.3379 (0.165) & \textbf{1406.32} &0.4821& 9.74\\  
    Boundary+TPS & \textbf{0.3340 (0.178)} & 1457.30 & 0.4830 & \textbf{9.42}\\ 
    Boundary+GR  & 0.3511 (0.183) & 1513.04 & \textbf{0.4833} &  \textbf{9.42}\\ 
    \bottomrule
  \end{tabular}
  \label{tab:Ablation1}
\end{table}

\begin{table}[!t]
    \scriptsize
  \centering
    \caption{Ablation studies on the decision choices of using different geometric elements of boundaries, text lines and vertical lines (V).}
  \vspace{-2mm}
  \begin{tabular}{lcccc}
    \toprule
    \textbf{Method} &  \textbf{CER}($std$) $\downarrow$ & \textbf{ED} $\downarrow$ &\textbf{MS-SSIM} $\uparrow$ & \textbf{LD} $\downarrow$\\
    \midrule
    DocUNet & 0.3955 (0.272) & 1684.34 & 0.4389 & 10.90\\
    Boundary & 0.3511 (0.183) & 1513.04 & 0.4833 & 9.42\\  
    Boundary+Text Line & 0.2081 (0.142) &908.84 & 0.4907 & 9.44\\ 
    Boundary+Text Line + V & \textbf{0.2068 (0.141)} & \textbf{896.48} & \textbf{0.4922} & \textbf{9.36}\\ 
    \bottomrule
  \end{tabular}
  \label{tab:Ablation2}
  \vspace{3mm}
\end{table}

We will analyze the feasibility of using a grid regularization optimizer to replace the traditional interpolation algorithm and the influence of the geometric information on the grid regularization optimizer.

\myparagraph{Grid Regularization \vs Traditional Interpolation.} In this ablation study, we use a grid regularization method to replace the traditional interpolation algorithm. All methods in Table~\ref{tab:Ablation1} use the same boundary points generated by our method. It demonstrates that our grid regularization optimizer can completely replace the traditional interpolation algorithm. In fact, $\phi$ obtained by our method can be regarded as two ruled surfaces of TFI with some distortion in accordance with the geometric structure of the text. And because we regard the grid regularization method as an optimization problem, the scalability is better than the traditional interpolation algorithm, which facilitates us to add geometric information into the subsequent algorithm.

\vspace{-1mm}
\myparagraph{The Influence of Different Geometric Information.}
In this study, we show the influence of boundaries, text lines and vertical lines for the final performance. As shown in Table~\ref{tab:Ablation2}, all those geometric elements are positive for document image dewarping.

\subsection{Limitations}
Although we have obtained inspiring results on the DocUNet benchmark dataset, there are still some issues with our approach. 
Firstly, due to the computational complexity of the optimization problem, the size of the deformation grid unit is currently set to be $128 \times 128$, which somewhat limits the density of the discrete geometric information and makes it difficult to continuously pick points in an extremely small area. This might be the main reason that the rectified result in the third column of Fig.~\ref{fig:result}~(f) has some parts moving toward the upper boundary. In order to get more precise rectified results, one can refine the grid by further pursuing some fast optimizers for the energy minimization task.  
Secondly, 
the current version of our method is not end-to-end, thus we may need to balance the different energy terms in the dewarping process.
Thus, one future direction is to develop an end-to-end trainable version of our method to automatically learn all the parameters.

\section{Conclusion}
\label{sec:concl}
This paper studies the problem of document image dewarping. By revisiting the deep-learning based dewarping approaches of document images, we address the remained challenge on the readability of the rectified document images in a geometric perspective with the text lines and image boundaries. With the learned geometric elements available, we design a grid regularization term on the deformation grid to estimate the 2D deformation field by solving an optimization problem. In our experiments, we demonstrate the effectiveness of our proposed approach with a new state-of-the-art performance on the DocUNet benchmark obtained.

\myparagraph{Acknowledgement.}
This work was supported by National Nature Science Foundation of China under grant 61922065, 62101390, 41820104006 and 61871299, China National Postdoctoral Program for Innovative Talents under grant BX20200248. This work was also supported by Alibaba Group through Alibaba Innovative Research (AIR) program.
The numerical calculations in this paper have been done on the supercomputing system in the Supercomputing Center of Wuhan University.
{\small
\bibliographystyle{ieee_fullname}
\bibliography{egbib}
}

\end{document}